# FuturICT – The Road towards Ethical ICT

Jeroen van den Hoven[1,a], Dirk Helbing[2], Dino Pedreschi[3], Josep Domingo-Ferrer[4], Fosca Gianotti[3], and Markus Christen[5]

[1] Philosophy Section, Delft University of Technology, Jaffalaan 5, P.O. Box 5015, 2600 GA Delft, The Netherlands
[2] Chair of Sociology, in particular of Modeling and Simulation, ETH Zurich, Clausiusstrasse 50, 8092 Zurich, Switzerland
[3] Dipartimento di Informatica, Universita di Pisa, Via Buonarroti 2 I-56125 Pisa, Italia
[4] Department of Computer Engineering and Maths, Universitat Rovira i Virgili, Av. Països Catalans, 26, E-43007 Tarragona, Catalonia
[5] Institute of Biomedical Ethics, University of Zurich, Pestalozzistrasse 24, 8032 Zurich, Switzerland & Psychology Department, University of Notre Dame, IN, USA

**Abstract.** The pervasive use of information and communication technology (ICT) in modern societies enables countless opportunities for individuals, institutions, businesses and scientists, but also raises difficult ethical and social problems. In particular, ICT helped to make societies more complex and thus harder to understand, which impedes social and political interventions to avoid harm and to increase the common good. To overcome this obstacle, the large-scale EU flagship proposal FuturICT intends to create a platform for accessing global human knowledge as a public good and instruments to increase our understanding of the information society by making use of ICT-based research. In this contribution, we outline the ethical justification for such an endeavor. We argue that the ethical issues raised by FuturICT research projects overlap substantially with many of the known ethical problems emerging from ICT use in general. By referring to the notion of Value Sensitive Design, we show for the example of privacy how this core value of responsible ICT can be protected in pursuing research in the framework of FuturICT. In addition, we discuss further ethical issues and outline the institutional design of FuturICT allowing to address them.

## 1 The Moral Case for FuturICT

Societies in the 21st century have reached a level of complexity that has made our understanding of them deeply problematic. Moreover, whatever provisionary understanding we manage to achieve is often outpaced by rapid change and new developments, making social and political interventions difficult. A major driving force of change is information and communication technology (ICT) that created new instruments for social exchange (e.g., social media, social network sites). Political decision makers, however, often have difficulties in keeping track of new media developments and in coming to grips with their dynamics and impact upon society and their vari-ous institutions like family, schools and universities, political parties, banks, business firms, or financial markets. World leaders are puzzled by the role of new media in political revolutions, such as the Arab Spring, social unrest such as the riots in London, or terrorist networks.

---
[a] e-mail: M.J.vandenHoven@tudelft.nl

Nevertheless, stakeholders affected by such interventions attribute responsibility to decision makers based on the extent to which they have attempted to make available the best and most comprehensive data and applied the best and most relevant models and theories to make sense of the data. This is based on the regulative ideal that we should avoid to intervene in a system that we do not sufficiently understand and of which we cannot reasonably predict how it will respond to our interventions. As long as this regulative ideal holds, the current situation creates an urgent need to increase our understanding of society. This results not only from the fact that uninformed and erroneous public policy in response to issues like terrorism, food safety, outbreaks of infectious diseases, financial crises, or environmental catastrophes, cost many citizens' lives and leads to enormous financial costs. It also is needed to preserve the trust and confidence of citizens in governments, as faltering policy may easily corrode government legitimacy.

The need for better understanding is underlined by the fact that many crises and disasters in modern societies are characterized by the lack of reliable real time data and action-oriented understanding of them. Examples are the 2010 Loveparade disaster in Duisburg/Germany[1], the EHEC[2] outbreak in several European countries in spring 2011, the terrorist outrage in July 2011 in Norway[3], and the riots in the UK in August 2011[4]. Such events routinely trigger the questions: Could we have known better? And: Should we have known better? These questions are appropriate, and unfortunately may even have a positive answer. For example, the work of Helbing and colleagues [23] (and other FuturICT researchers) shows that we can gain a better understanding of crowd disasters, helping to improve the design of similar events. The ethical conclusion is that a responsible organisation and design of future Loveparades requires due attention to the lessons learned and the science of the relevant phenomena. This example also shows that not only the generation of data and scientific understanding is a major task, but also the transfer of this knowledge to deciders and players in the real world.

Since ICT has become constitutive of many social phenomena, there is no way in hoping to understand the current information society and human behavior within this society without the technology and sciences relevant to the study of computer systems and networks. This is where the basic idea of FuturICT sets in: FuturICT is a community of scientists, users and citizens that is ready to jointly take the responsibility to shape a global knowledge resource that will provide on-line functionality and social networking platforms allowing for deeper individual and collective understanding of modern society and its problems. This endeavor, which is in more detail described by other contributions of this special issue (for an overview see [21]), is based on four main ethical postulates:

1. *Epistemic Responsibility:* Those who bear responsibility for policies and interventions in complex systems have a responsibility for creating the knowledge conditions which allow them to do the best they can. Decision makers are framed by a given epistemic context and are dependent on the information infrastructure put at their disposal. The quality of their decisions and judgments is in many cases determined by the quality of their knowledge tools (i.e., information systems, programs and data). Responsibility of decision makers therefore importantly concerns the design *ex ante* of epistemic resources and information infrastructures, which is a major aim of FuturICT.

---

[1] In a human stampede 21 visitors died and 541 were injured.
[2] Enterohaemorrhagic E. coli (EHEC) is a bacterium that causes severe foodborne disease.
[3] A single assassin killed 77 persons in a combined bomb attack and shooting.
[4] The London Police during the riots monitored Blackberry Messenger and Twitter realtime to prevent planned attacks. They were struggling to keep ahead of the rioting crowds and also considered closing off access to social networks (see discussion in [9]).

2. *Social Knowledge as a Public Good:* A broad range of information about society ought to be accessible to all citizens under conditions of equal opportunity. FuturICT forms a counter-balance against the buildup of information monopolies in important domains in society by private sector companies and thus contributes to a just and fair information society.
3. *Privacy by Design:* Privacy is an essential moral constraint for achieving knowledge and understanding of social reality in information societies. Although the term refers to a broad range of moral rights, needs, claims, interests, and responsibilities concerning (information about) the person, personal lives, and personal identity, privacy is essential for the flourishing of individual human beings. Data protection technology needs to be developed in tandem with data mining techniques and E-social science. The development of new forms of Privacy by Design is a central objective of FuturICT.
4. *Preserving Trust in Information Society:* Trust implies a moral relationship between the truster and the trustee, a relationship that is partly constituted by a belief or an assumption that the trustee will act from the moral point of view. In complex ICT-shaped environments trust requires that those in charge of the design of the environment, in which the trust relationship is situated, are as explicit and transparent as possible about the values, principles and policies that have guided them in design. This is a fourth guiding principle for FuturICT, whose ultimate goal is the fair information society, where there is openness and transparency about the values, principles and policies that shape it.

By building on these postulates, we explore the ethical aspects, issues and problems of FuturICT as a comprehensive source of social knowledge. We will show that the ethical problems raised by the pervasiveness of ICT in the information society, which FuturICT aims to resolve, are to a large extent co-extensive with the problems that FuturICT may encounter as a research project: problems of privacy, accountability, liability, access, property, freedom, and democracy. In doing so, we proceed as follows: In the next Section, we briefly sketch how ICT currently transforms traditional social sciences and humanities disciplines and in this way we provide the ground for FuturICT. In Section 3 we describe how the principle of Value Sensitive Design was developed in the context of ICT as a guiding principle to make information technology sensitive for social and ethical issues. In Sections 4 to 6 we outline in detail why privacy is a core value for the information society, how it is endangered, and how privacy will be protected in the course of advancing research in the framework of FuturICT. In Section 7 we discuss other ethical issues that will have to be treated within FuturICT, and in Section 8 we sketch the institutional design of this research project in order to be able to integrate these issues. The concluding Section 9 summarizes the main points made in this paper.

## 2 The Rise of E-Social Science and E-Humanities

Under the labels of 'E-social sciences' and 'E-humanities' (or: 'digital humanities') traditional disciplines are making first steps at the beginning of the 21st century to provide ICT-based insights into the information society. Europe's rich and variegated intellectual traditions provide fertile soil to do so. Europe not only gave birth to natural science in our current understanding, it also was the birthplace of social science, economics, and moral philosophy. In recent times, increasing interest in a new and computer supported exploration of the unity of what was erstwhile referred to as "the moral sciences" (the study of man and society) is detectable. These computer supported "moral sciences" are provoked first by the rise of information technology and are somehow parallel to developments in other fields like elementary particle physics, astrophysics, genetics, bio-informatics, and climatology. The availability of digital computers, large amounts of data, and the application of mathematical models in search of recurrent patterns are seen in many disciplines as being of equal value compared to the traditional empirical cycle of hypothesis formulation, observation, testing, validation or falsification, and hypothesis reformulation. Furthermore, this development is provoked by the application of formal and computer assisted methods to supplement qualitative, hermeneutic, phenomenological and ideographic methods in the social sciences and humanities. Theories and methods emerging from fields like behavioral and computational economics, game theory, social choice theory, rational choice theory, mechanism design, social software, principle agent theory, agent based modeling, or modal logics are now applied to the social, moral, legal and economical realms. Information technology has brought fascinating new knowledge and has shifted the social sciences and humanities into a new phase.[5] Also ethics has made a sharp empirical turn in the last decade, and methods and tools emerging from psychology, behavioral economics, computer science, cognitive/ social neuroscience and social science are increasingly acknowledged [8].

This development has also real-world consequences: ICT – in the form of (agent based) simulations – is brought to bear upon institutional design and public policy design. Social interaction platforms or political and economic institutions like auctions (e.g., of emission rights or radio frequencies), voting schemes, taxation regimes, markets, or pension schemes, are now designed on the basis of new knowledge gained in these new fields, and computer simulations often serve as a sandbox. This type of E-social science research and E-humanities is only in its infancy – and FuturICT promises to make a formidable step in the direction of a better understanding of social complexity, scale and dynamics that we can only disregard at our own peril.

Certainly, this turn to E-social science and digital humanities, often captured by the slogan 'Big Data', has also been criticized by various scholars. We mention only one example: Microsoft researcher danah boyd [sic] brings forward six 'provocations' to spark the debate about issues of Big Data [6]. First she points out that the automation of research changes the definition of knowledge. Big Data refers to a "computational turn in thought and research" causing a radical shift in how we think about research. Not only the depth and scale of the capacity to collect and analyze data is without precedent, Big Data also instigates "a profound change at the levels of epistemology and ethics. It reframes key questions about the constitution of knowledge, the processes of research, how we should engage with information, and the nature and the categorization of reality" ([6], p. 3).

Second, claims to objectivity and accuracy accompanying Big Data may often be misleading. Although it is alleged that Big Data offer humanistic disciplines and social sciences an objective method, "[...] in reality, working with Big Data is still subjective, and what it quantifies does not necessarily have a closer claim on objective truth [...]" ([6], p. 4). In designing a data model and interpreting outcomes researchers are involved and are thus "based on subjective observations and choices." Moreover, data sets can be "unreliable, prone to outages and losses, and these errors and gaps are magnified when multiple data sets are used together" ([6], p. 5). These limitations and biases should be taken into account when analyzing Big Data. Third, "bigger data" are not always better data. The presumption holds sway that bigger is better and as a result "the core methodological issues in the social sciences are no longer relevant" ([6], p. 6). Twitter, for example, "has become a popular source for mining Big Data, but working with Twitter data has serious methodological challenges that are rarely addressed by those who embrace it" ([6], p. 7). Furthermore, when datasets from different sources are combined, ethical issues such as privacy and problems of accountability may be magnified. Also for some research questions a smaller data set is more appropriate than a big one. Fourth, combining different data sets to generate bigger data, often involves taking data out of their original context: "When two datasets can be modeled in a similar way, this does not mean that they are equivalent or can be analysed in the same way" ([6], p. 8).

---

[5] Examples that highlight this development are the National Centre for e-Social Science in the UK (http://www.ncess.ac.uk), and the Digital Humanities Observatory in Ireland (http://dho.ie).

Fifth, the fact that some data are accessible does not entail that it is morally right to access them. Researchers have a responsibility to deal with data ethically, even when data are publicly accessible, especially when it is impossible to obtain consent by each person represented in the data as is often the case with Big Data. This means "both accountability to the field of research, and accountability to the research subjects" ([6], p. 11).

Finally, limited access to Big Data creates new digital divides. For example, access can be limited to those who own the data and are not willing to share, only to those who are willing and able to pay high prices for data (for instance high tier universities as opposed to lower ones), or to those who have the computer skills needed to access and analyze the data in a technical sense. This creates divides between those who have access and those who do not, restricting possibilities to do research and to reproduce or evaluate the methodological claims made by researcher who have access. Furthermore, "the difficulty and expense of gaining access to Big Data produces a restricted culture of research findings" ([6], p. 13). Suppliers of data may restrict access to those who do not choose questions that are contentious to the practices of those suppliers.

These observations are all pertinent both to the future of ICT as well as to scientific disciplines that heavily rely on ICT. With respect to the latter, the second, third and fourth observation pertain to methodology of data mining in the social sciences, and the other observations pertain to issues like data protection, epistemic responsibility and distributive justice that will be discussed later in this contribution.

## 3 Value Sensitive Design

The development of ICT can be described in many ways. One way of characterizing it is that ICT has opened up to its environment: to users, to organizational and social context, and to society at large. In the first phase of its development in the 1960s and 1970s ICT was largely the outcome of a technology push, centered around core computational functionality. Hardly anyone bothered to ask about users, use and usability. Digital computers were fascinating technology, solutions looking for problems. In the second stage of the development in the 1980s and 1990s – after many failed ICT projects – one gradually started to realize that out there were human ICT users of flesh and blood, real-world organizations, and a society. Thus, it would be wise to accommodate user requirements already in the early stage of application development. But this was still a very minimal way of taking the needs and wishes of users, organizations, and society into account, namely as mere constraints on the successful implementation of systems. Now, we are entering a third phase in ICT development, where the needs of human users, the values of citizens, and the big societal questions are in part driving research and development in ICT. A good example of this "social and value" turn is the Californian Institute CITRIS (Centre for IT Research in the Interest of Society; see `http://www.citris.berkeley.edu`). In this phase, social and moral values are no longer seen simply as "risk factors" or constraints, but also as drivers of innovation. It brings within reach the idea that applications are developed in order to serve and support values and serve the interest of society.

The idea of making social and moral values central to the development of ICT and to embed values in ICT artifacts (like architecture, code, interface, integrity constraints, ontologies, authorization matrices, or identity management tools) has been advanced by the Stanford Computer Science Department (e.g. by Terry Winograd). It is now referred to as Value Sensitive Design (VSD). In the VSD approach the focus is on incorporating moral values – as a type of non-functional requirements – into the design of technical artifacts and systems. In doing so one takes a look at software engineering design from an ethical perspective. One can intentionally and explicitly design for values such as accountability, safety, inclusion, privacy, trust, or sustainability [15,16]. In VSD the incorporation of moral values into the design is a primary goal instead of a by-product. As has been argued by Van der Hoven [46], VSD is at the same time a way of doing ethics that aims at making moral values part of technological design, research and development. However, the VSD approach can only be used if we manage to be explicit about the variety of moral reasons for desirable features of systems, can formulate them as "non-functional requirements", and have at our disposal a transparent way of decomposing them into more detailed functional requirements. We will exemplify this point for privacy in Section 6.

Value Sensitive Design also helps us to look more specifically at ways of reconciling different and opposing values in engineering design [47]. This idea can be illustrated as follows. As a society we value privacy, but at the same time we value security and the public availability of information about citizens in order to be able to effectively fight crime. The pursuit of these values creates a tension which is exemplified e.g. in the debates about Closed Circuit TV cameras in public places.[6] We either hang cameras in all potentially risky spots in the public space and thereby create the desired level of security, but give up on our privacy. Or we respect privacy and refuse to hang cameras excessively, but settle for less security. Ideally we want both privacy and security – and technology may actually provide a solution for this dilemma. Smart camera systems may allow us to enjoy the functionality the technology can offer, and at the same time respect the moral constraints on the availability of personal data that privacy requires. Smart, privacy enhancing technology (an example of Privacy by Design, see Section 6) may allow us to stipulate who gets access to which recordings, on which conditions, how long the images are stored, how they may be used and merged with other databases. Thus, innovations or "smart" technologies often manage to reconcile previously irreconcilable values or preferences by design.

The High Level Advisory Group on ICT to the European Commission (ISTAG) has in its 2020 vision report of 2008 drawn attention to the appropriateness and suitability of a Value Sensitive Design approach for Europe's thinking on responsible innovation. Also the Ethics Department of the Directorate-General for Research and Innovation of the European Commission is thinking along Value Sensitive Design lines in drafting their Ethics Program towards the EU Horizon 2020 Program. Responsible Innovation according to the envisaged Ethics program of the EU is not so much about enhancing "the self parking car" and other convenience enhancing gadgets, but about designing for the smart resolution of the crucial tensions in society along the lines sketched above.

---

[6] Closed-circuit television (CCTV) is the use of video cameras to transmit a signal to a specific place, on a limited set of monitors. Surveillance of the public using CCTV is common in many areas around the world including the UK, where there are reportedly more cameras per person than in any other country in the world.

Given these considerations, also FuturICT is committed to the idea of Value Sensitive Design with respect to responsible innovation in data mining and fair practices of knowledge discovery in databases. One core value for FuturICT is privacy – and the main idea is that researchers in FuturICT study and develop tools, algorithms, and architectures that should express the value of privacy. This conception of Privacy by Design, its moral justification and the threats that made privacy vulnerable in the information society, will now be discussed in more detail in the next three Sections.

## 4 The Moral Importance of Privacy and Data Protection

'Privacy' is a cluster concept that unifies a number of moral considerations in support of data protection: (a) the prevention of harm to data-subjects, (b) "equality of arms" and fairness in markets for personal data to prevent economic exploitation of citizens, (c) informational justice and non-discrimination, (d) respect for moral autonomy, and (e) the connection between privacy, creativity and innovation. As privacy is a central value endangered by various ICT-driven developments[7] and as also the research promoted within FuturICT has the potential to undermine it [18], privacy is the main focus of our contribution. In the following, we provide a discussion of the moral reasons for privacy protection, whereas the next Section is devoted to privacy threats, and in Section 6 we outline the concept of Privacy by Design, as well as some of the tools and design principles that express privacy and data protection constraints and that will be implemented in pursuing research in the framework of FuturICT.

When it comes to private data, some people suggest that privacy is mainly in the interest of dishonest, criminal and deviant persons, who have an interest to hide aspects of their lives from public view. This is a dangerous misconception. Privacy has been granted not as a concession of the state to the individual, but because there are very strong moral reasons to allow human beings to exercise control over their personal information. Human beings need it to flourish – and modern societies thus need data protection in order to flourish.

Information technology allows us to generate, store and process huge quantities of data. Search engines, satellites, sensor networks, scientists, security agencies, marketers, database managers are processing terabytes of data per day. A good part of these data are about persons – about their characteristics, their thoughts, their movements, behavior, communications and preferences – or they can be used to produce such data[8]. All countries and cultures in the present and past have constrained access to certain types of personal data in some way or the other [31]. There are etiquettes, customs, artifacts, technologies, laws as well as combinations thereof, which prevent or proscribe the use or dissemination of personal information. Walls, curtains, doors, veils, sealed envelopes, sunglasses, clothes, locked cabinets, secure databases, cryptographic mechanisms, passwords, . . . – all these instruments serve the purpose of preventing individual persons to acquire and use information about other persons. Privacy issues are often discussed in the context of a specified social sector or professional domain such as health care, social security, homeland security, search engines, marketing, or policing. And it may also include issues like access to scans and digital images of our brains, or tracking and tracing RFID tagged personal belongings, everyday objects and consumer products.

---

[7] See for a recent overview of the state of the art of data protection in Europe and some of the challenges it faces: [20]

[8] There is a widely accepted convention to distinguish between data (raw data), information (meaningful data) and knowledge. Our main concern here is with data or the raw material that can be utilized and interpreted by a variety of methods and tools to serve many different purposes. We here explore the moral foundations of data protection. In many cases not much depends on whether we use "data" instead of "information".

Given this large spectrum of issues, privacy is construed in various ways: as a need, a right, a condition, or an aspect of human dignity. Sometimes it is construed as intrinsically valuable, sometimes it is construed as deriving its value from other sources, e.g. from the fact that it is conducive to autonomy or freedom (see [10,33,34,37,40,49]). The biggest part of privacy research is concerned with the moral justification of a right to privacy, but there is little agreement among experts in that respect[9]. Nevertheless, the fact that privacy is important is undisputed among privacy scholars, although the concept is vague, fuzzy and hard to explicate or pin down [10], and there are a number of fundamental moral reasons for engaging in serious data protection. These moral reasons provide the grounds to have principles like those of the *EU95 Data Protection Act*,[10], or the OECD principles[11]. All these legal regimes have as their normative core the autonomy or informational self determination and the right to control their personal data as expressed by the crucial requirement of informed consent. In the following, we outline five bundles of reasons in favor of constraining the flow of personal data or identity relevant data.[12]

### 4.1 Information Based Harm

The first type of moral reasons for ensuring privacy is concerned with the prevention of harm. Information is a very useful thing to have, for every benevolent human person – but also for criminals, crooks and people with bad intentions. Thus, aggregates of Big Data are a primary target for cyber-criminals, which provides a big challenge for security experts. The 2011 Symantec Internet Security Threat Report mentions a steep increase in the number of new malicious programs (malware) identified (more than 240 million in 2009, 286 million in 2010). It estimates the number of internet users (companies and individuals), who have been victims of cyber-attacks trying to steal money or confidential information, to be of the order of 360 millions. More and more attacks aim at identity theft. Sixty percent of all data breaches that revealed identities were in fact the result of hacking. An incomplete list of internet risks include: theft of passwords, PIN-codes, data and personal information (identity theft) by the use of viruses, worms, and Trojan horses; data manipulation (e.g., to create wrong 9evidence); attempts at blackmail by using ransomware[13]; and "information pollution" like damaging rumors or spam.

---

[9] See for a recent and comprehensive discussion of philosophical accounts Beate Roessler [37]. Already in 1986 Thomas Perry indicated that privacy was a battleground for different positions in applied ethics [35].

[10] Directive 95/46/EC is the reference text, at European level, on the protection of personal data (available at: http://europa.eu/legislation_summaries/information_society/data_protection/l14012_en.htm). Revised directives are currently drafted and will come into force in Europe in the next decade.

[11] The *OECD Guidelines on the Protection of Privacy and Transborder Flows of Personal Data* are available at: http://www.oecd.org/internet/interneteconomy/oecdguidelinesontheprotectionofprivacyandtransborderflowsofpersonaldata.htm

[12] We need a broader category, because there are data that would now fall outside of the scope of the legal definition of "personal data", that are still relevant to the determination of a person's identity or his or her (re)identification in the future. An argument for a broader notion of personal data is discussed in Van den Hoven [48].

These threats may all cause serious harm to individuals, but they also seriously undermine the trust of people in the internet and services based on it. For example, the theft of access data for electronic banking through phishing attacks has recently become a wide-spread problem. However, trust is essential for economic exchange. Systems which would not work effectively without a certain level of trust include: e-mail, e-Banking, e-Business, e-Governance, and social networking. As the number of services based on the internet increased substantially in the information society, we are confronted with a new vulnerability to harm on the basis of personal data-theft, identity fraud and the like. We could refer to this as 'information based harm', whereas the term 'harm' is a broad category and covers different types of harm such as reputational harm, or what Anthony Appiah calls 'probabilistic harm' (the harming of people by decreasing their chances of getting some good) [2].

To summarize, in an information society, persons can be harmed in a great number of ways on the basis of the information that is available on them. Constraining the freedom to access information of persons can thus be justified on the basis of the Harm Principle (the actions of individuals should only be limited to prevent harm to other individuals) proposed by John Stuart Mill in *On Liberty*. Protecting information that identifies and characterizes individuals, diminishes the freedom of everyone to know, but also diminishes the likelihood that individuals or groups will be harmed.

**4.2 Informational Inequality**

The second type of moral reasons for ensuring privacy is concerned with equality and fairness. More and more people are keenly aware of the benefits the market for identity information can provide. If a consumer buys coffee at a modern shopping mall using, e.g., a rebate card, information about that transaction is generated and added to his file or profile. Many consumers now begin to realize that every time they buy something, they can also sell something: the information about their purchase or transaction, the so-called transactional data. Likewise, sharing information about ourselves on the internet when visiting Web sites, using browsers, or interacting with software agents may pay off in terms of more and more adequate search results, purchasing recommendation, discounts, or convenience. Many privacy concerns have therefore been – and will be – resolved in *quid pro quo* practices and private contracts about the use and secondary use of personal data. But although a market mechanism for trading personal data seems to be kicking in on a global scale, not all individual consumers are aware of their economic opportunities. And if they are, they are not always in a position to trade their data or pursue their interests in a transparent and fair market environment, so as to get a fair price for them and avoid becoming an easy target of price discrimination.

Consumers do not always know what the implications are of what they are consenting to when they sign a contract for the use of identity relevant information. We simply cannot assume that the conditions of the developing market for identity related information guarantee fair transactions by independent standards. Moral and legal constraints for the flow of personal data based on principles of fairness need therefore to be put in place in order to guarantee "equality of arms", transparency and a fair market for identity relevant information as a new commodity.

---

[13] Ransomware is a class of malware which restricts access to the computer system that it infects, and demands a ransom paid to the creator of the malware in order for the restriction to be removed.

### 4.3 Informational Injustice and Discrimination

A third and very important moral reason with respect to privacy is concerned with justice in a sense which is associated with the work of Princeton philosopher Michael Walzer. He has objected to the simplicity of Rawls' conception of primary goods and universal rules of distributive justice by pointing out that "there is no set of basic goods across all moral and material worlds, or they would have to be so abstract that they would be of little use in thinking about particular distributions" ([50], p. 8.). Goods have no natural meaning. Their meaning is the result of socio-cultural construction and interpretation. In order to determine what a just distribution of the good is, we have to determine what it means to those for whom something is a good. In the medical, the political, or the commercial sphere, there are different goods (e.g., medical treatment, political office, money) which are allocated by means of different allocation criteria or distributive practices. Medical treatment is allocated on the basis of need, political office on the basis democratic election, and money on the basis of free exchange. What ought to be prevented – and often is prevented as a matter of fact – is the dominance of particular goods across spheres. Walzer calls a good dominant if the individuals who have it can command a wide range of other goods due to this possession of this good. A monopoly is a way of controlling certain social goods in order to exploit their dominance. In that case, advantages in one sphere can be converted in advantages in other spheres. For example, this happens when money (commercial sphere) can buy you a vote (political sphere), would give you preferential treatment in healthcare (medical sphere), or would get you a university degree (educational sphere). We resist the dominance of money – and other social goods (land, physical strength) – for that matter and we think that political arrangements allowing for it are unjust. No social good X should be distributed to men and women who possess some other good Y merely because they possess Y and without regard to the meaning of X.

What is especially offensive to our sense of justice is the allocation of goods internal to sphere A on the basis of the distributive logic associated with sphere B, the transfer of goods across the boundaries of separate spheres, and the dominance and tyranny of some goods over others. In order to prevent this, the 'art of separation' of spheres has to be practiced and 'blocked exchanges' between them have to be put in place. If the art of separation is practiced effectively and the autonomy of the spheres of justice is guaranteed then "complex equality" is established. One's status in terms of the holdings and properties in one sphere are irrelevant – *ceteris paribus* – to the distribution of the goods internal to another sphere.

Walzer's analysis also applies to information. The meaning and value of specified information is local; and allocation schemes and local practices that distribute access to information should accommodate those local meanings and should therefore be associated with specific spheres. For example, many people do not object to the use of their personal medical data for medical purposes (i.e., are confined to the medical sphere), whether these are directly related to their own personal health affairs, to those of their family, and perhaps even to their community, or the world population at large – as long as they are absolutely certain that the only use that is made of this information is medical, i.e. to cure people from diseases. However (set aside some well-defined exceptions), they do object to their medical data being used to classify them or disadvantage them socio-economically, to discriminate against them in the workplace, to refuse them commercial services, to deny them social benefits, or to turn them down for mortgages or political office on the basis of their medical records.

These considerations lead to a third moral reason to constrain actions regarding identity relevant information: prevention of informational injustice, that is, disrespect for the boundaries of what we may refer to, following Michael Walzer, as 'spheres of justice' or 'spheres of access'. Thus, what is often seen as a violation of privacy, is oftentimes more adequately construed as the morally inappropriate transfer of personal data across the boundaries of what we intuitively think of as separate 'spheres of justice' or 'spheres of access' [43,48]. This complex moral reason also accounts for the moral wrongness of discrimination. Discrimination of a person P in a particular context implies the use of information about P to her disadvantage while the information is morally irrelevant to that context. The art of separation of spheres (or contexts of use) and the blocked exchanges (Walzer, Van den Hoven) would prevent discrimination. It secures that information is used in contexts where it is relevant and morally appropriate.[14]

### 4.4 Moral Autonomy

A fourth type of moral reasons could be referred to as moral autonomy, i.e. the capacity to shape our own moral biographies, to present ourselves in a way that fits our self-understanding, to reflect on our moral careers, and to evaluate and identify with our moral choices without the critical gaze and interference of others and without a pressure to conform to the 'normal' or socially desired identities. We want to be able to present ourselves and be identified as the ones we identify ourselves with. David Velleman, in his analysis of shame and privacy, draws attention to self-presentation as a constitutive feature of moral persons, namely their capacity and need for self-presentation. What it means to be a person is according to Velleman, to be engaged in self-presentation. Persons "have a fundamental interest in being recognized as a self-presenting creature" [49]. On occasions where we fail as self-presenters, we feel that our privacy is lost and we typically experience shame. The realm of privacy according to Velleman is therefore the central arena for threats to one's standing as a social agent.

A moral person is thus characteristically engaged in self-presentation, but experiences the normative pressures that public opinion and moral judgments of others impose. When information about a person P becomes available, it facilitates the formation of beliefs and judgments about P. They may induce him to behave and feel differently than he would have done without them. This is what Berlin calls "the most heteronomous condition imaginable"[15]. What others know about you can radically affect your view of yourself, although seeing yourself in a way others see you, does not necessarily make your view of yourself more true or more adequate ([4], p. 277 resp. p. 288).

Stereotyping is an extreme case of casting people and pre-empting their choice to present themselves. To modern individuals, who have cast aside the idea of fate, who live in a highly volatile socio-economic environment, and who confront a great diversity of audiences in different roles and settings, the rigging of one's moral identity by means of public opinion, beliefs and judgments of others is felt as an obstacle to "experiments in living", as Mill called them. The modern individual wants to be able to determine himself morally or to undo his previous determinations, on the basis of more profuse experiences in life, or additional factual information. Some have argued that privacy creates a time out from social morality, in order to engage in ever new experiments in living. Privacy covers purely self-regarding acts and therefore implies a right to non-justification [30].

---

[14] Helen Nissenbaum [34] has introduced the term "contextual integrity" to refer to this idea.

[15] "I cannot ignore the attitude of others with Byronic disdain, (…) for I am in my own eyes as others see me, I identify myself with the point of view of my milieu." [5], p. 156)

The conception of the person as being morally autonomous, as being the author and experimenter of his or her own "moral career", provides a justification for constraining others in their attempts to engineer and directly or indirectly shape the subject's identity by stereotyping or using identity management tools and techniques. Data-protection laws thus provide protection against the fixation of one's moral identity by others than oneself. They do so by requiring informed consent for the processing of identity relevant information.

### 4.5 Freedom, Creativity and Innovation

There is one additional point to be made about privacy, namely its connection to creativity and innovation. This builds on the moral autonomy argument for data protection, the ethical value of creativity, and our current understanding of the emergence of new ideas. Innovation usually starts off in a minority position, and there are, e.g., usually only a few supporters of a new idea or only a few customers of a new product. In other words, there is little innovation without the existence of minorities. As is known from evolutionary theory, innovation thrives best when there is a large diversity of variants. In other words, diversity or pluralism is the motor that drives innovation. Would the individual be completely conformist with respect to the own thinking and acting – and the pressure for conformity is high, as for example the famous Asch experiments have demonstrated[16] –, the innovation rate and, as a result, the adaptability of a society to changing (environmental) conditions would be poor. One may speculate that this is a core reason why totalitarian regimes are sooner or later destined to fail.

Privacy is an instrument to protect minorities and enable pluralism; and lack of privacy and the resulting conformity forms a "Tyranny more formidable" (Mill), and may lead to stifling political correctness, echoing and cascading of false beliefs. Societies may lose resources of criticism, creativity and the wellsprings of innovation. Without privacy, pluralism is in danger. Social imitation created herding effects, which are often misleading. For example, the financial crisis starting in 2008 may have been caused to a large degree by such herding effects, which led to extremely expensive mistakes. Herding-related mistakes would become even more likely, when people are put under pressure to conform with majority opinions or behaviors. Actually, many of the current web services and recommender systems, which reinforce dominating opinions, may support such unwanted developments.

## 5 The Ever New Faces of Privacy Vulnerability

Privacy is, as we have outlined, a core valued backed by various moral reasons. But in the information society, privacy has become more and more vulnerable. Privacy can be endangered by camera surveillance, monitoring of internet communications, retention of internet traffic data, disclosure of passenger lists to security agencies, availability of individual medical information in the public health system, linking and matching of databases in the social security to detect fraud, or sifting and trawling through financial databases in order to find suspect transactions. Citizens may become the target of strategies of big data mining companies that are driven by market competition and the logic of profit, they may be subjected to extensive government surveillance schemes, or they may become victims of hacktivists or organized crime. Data mining techniques improve every day, while regulations and control tools possibilities are lagging far behind. For example, tracking the source of collected information – once it is stored in secured and not publicly accessible databases –, or knowing who has access to which kind of personal data is virtually impossible.

---

[16] In the experiment, individuals gave predominantly wrong answers – against their own judgment –, if the people before them did so as well.

Even without a sophisticated and expensive data mining system, criminals can collect illicit data easily through web browsers, as these are daily affected by new malicious exploits. One example of such an attack is a technique called "history stealing", by which personal surfing habits of internet users are extracted. Scientific literature on the topic is vast, and a study conducted on more than 270,000 users found that 76% of them were vulnerable to history detection by malicious Web sites. More recent browsers such as Safari and Chrome were even more affected, with 82% and 94% of vulnerable users [26]. Unfortunately, there is yet another privacy issue related to recent generations web browsers: their inherently high customization capabilities have made them unique, and therefore traceable. In fact, even if disabling cookies, and blocking history stealing-like exploits, individual web surfing can still be reconstructed by simply following the customized "fingerprint", which the browser is carrying around from site to site. This fingerprint is actually made up by all the configuration information that the browser is exposing to remote Web sites. According to the Electronic Frontier Foundation, information such as which plug-ins are installed, which fonts are available and which operating system the browser is running on, can create a unique portrait of 94% of the visitors [14].

Unethical or dishonest intent is not the only pitfall glooming over on-line data sharing. Even in a scenario in which one has consciously provided his or her own personal data to a company that is using them lawfully, unforeseen issues can suddenly arise. For example, such a company could be sold or merged with another one, or simply, the data could be sold, based on a change in the data handling policies. Users are typically not notified of such changes, and they usually have no effective possibility to draw back their data and their consent to use them. In fact, because of the continuous updating and modification of the terms of use, the Electronic Privacy Information Center (EPIC: http://epic.org) has filed a formal complaint at the US Federal Trade Commission. In 2010, US senator Charles Schumer (D-N.Y.) has petitioned the Federal Trade Commission to request that the agency address the issue of social network privacy policies.[17] Moreover, some national data protection commissioners have in 2009 and 2010 publicly warned of using certain social network sites.[18] Also in 2010, the vulnerability of these services has been demonstrated by a security consultant, who downloaded 100 million user profiles and made them publicly available for download.[19]

Joining groups within social networks can offer another exploit for potentially malicious de-anonymization attacks. A recent paper showed that 42% of users that utilize groups can be uniquely identified [51]. These results are noteworthy, because traditional privacy attacks were based on aggregating information from multiple datasets. Such methods were based on collaborative filtering and enabled an efficient and highly reliable characterization of a person from a few data. The underlying technology is quickly advancing, and it may give service providers, such as mobile phone, internet television, or social gaming centers an unprecedented amount of personal informa-

---

[17] News report: Senator calls on FTC to tackle social-net privacy; (http://news.cnet.com/8301-13577_3-20003415-36.html).

[18] News reports: EU warns on Facebook privacy (http://www.nytimes.com/2009/01/27/technology/27iht-facebook.4-417144.html); German minister warns Facebook over privacy rules (http://blog.foreignpolicy.com/posts/2010/04/05/german_minister_warns_facebook_over_privacy_rules)

[19] News report: Details of 100 m Facebook users collected and published (http://www.bbc.co.uk/news/technology-10796584).

14tion. Other risks for the privacy of users emerge, when companies are forced to reveal private data to governments or legal institutions. In addition, when data are not subpoenaed or stolen by cyber-criminals outside of the company, they can be leaked in the most fanciful ways, which go from mislaying a physical device containing sensitive information, to the dishonest action of a single employee from inside the company. Finally, the example of WikiLeaks, a Web site that publishes secret information, news leaks, and classified media from anonymous news sources and whistleblowers, shows that privacy violations may even be justified by referring to values like transparency, although the ethical evaluation of WikiLeaks is ambivalent [32].

## 6 Adressing the Issue: Privacy by Design

So far, we have outlined the moral reasons for privacy protection and the various threats which make this value vulnerable in the information society. Given this setting, part of the solution to privacy problems is that the technology itself can take care of compliance. There are roughly two ways in which ICT can come to the aid of the protection of privacy and other values such as fairness and equality: first by helping us to shift from contexts where moral and legal prohibitions can be circumvented, to an enhanced context where this is no longer possible (*ex ante*), and second, where superior techniques of detection of non-compliance are in place (*ex post*). This is for example achieved by discrimination detection algorithms or data mining techniques to identify anti-trust violations, corruption, kick-back payments and tax fraud [38,39].[20] The framework to advance is thinking is Privacy by Design, an exemplification of the Value Sensitive Design approach (Section 3). The basic idea is to inscribe privacy protection into the analytical technology by design and construction, so that the analysis takes the privacy requirements in consideration from the very start. Privacy by Design [21] is a paradigm that promises a quality leap in the conflict between data protection and data utility. It requires articulation of principles, specification and decomposition of moral requirements. This supplements the functional requirements for design of FuturICT. In doing so, one has also to take into account that privacy concerns can cause serious obstacles to socio-economic data mining, while in many cases such data mining would be in the public interest, e.g. when gaining a better understanding of socio-economic problems, how they arise and how they can be addressed. Thus, the task is to find a privacy-respecting way of socio-economic data mining without making this endeavor impossible [28,29].[22]

FuturICT plans to develop a platform for accessing global human knowledge as an open resource and a public good. This ought to implement two-sided transparency between the knowledge access system and its users. On one side, the methods used to access knowledge have to be made publicly available to the purpose of full comprehension, assessment of biases and replicability of results. On the other side, the users (either individual persons or institutions) will identify themselves and specify their purposes of knowledge access; their requests and the obtained answers will be recorded and made available for audits. Two-sided transparency is required by and fosters two-sided (mutual) responsibility: all parties have incentives to be accountable for the quality and purpose of their interactions and exchange. Complex network theory and data mining can help us understand and measure these effects, and help to discover how proper designs that take into account both the properties of ICTs and social behavior can exploit the network effect and the large numbers to minimize the probability of (1) disruptive hacking pursued by isolated criminals, (2) illegitimate commercial use by corporate parties, and (3) illegitimate surveillance of state and non state actors.

---

[20] The neural net and bot technology used by Wikipedia to moderate an essentially open community and to detect malicious activity is a good example of what can be achieved [11].

[21] The European Data Protection Supervisor Peter Hustinx is a staunch defender of the Privacy by Design approach and has recommended it as the standard approach to data protection for the EU, see http://www.edps.europa.eu/EDPSWEB/webdav/site/mySite/shared/Documents/Consultation/Opinions/2010/10-03-19_Trust_Information_Society_EN.pdf. Ann Cavoukian – Privacy Commissioner of Ontario, Canada – has been one of the early defenders of the Privacy by Design approach. See the principles that have been formulated http://privacybydesign.ca/about/principles/

[22] For an in-dept analysis of internet research ethics, see the entry in the *Stanford Encyclopedia of Philosophy*, available at: http://plato.stanford.edu/entries/ethics-internet-research/.

In such a mutually-responsible interaction, the system knows a lot about you. You are exposing your profile and data, but this feature can also be used to your advantage (sometimes referred to as "sous-veillance"), e.g., the system can be used for fraud detection and exposure of illegitimate usage. However, the human activity data that are sensed and analyzed within a techno-social system contain very sensitive information that can be traced back to individuals, possibly intruding into their privacy. For this reason, the methods that expose human related knowledge ought to be privacy-preserving; once put into operation in the open knowledge access system, they must perform with suitable transformations/samples/aggregations of the human data/patterns/models, so as to achieve the desired anonymity of the individuals whose data was collected in the process. Users within the FuturICT information sphere will be provided with suitable and morally justified incentives to comply with established and recognized data protection requirements. On the basis of the Value Sensitive Design approach and the more specific vantage point of Privacy by Design we have provisionally arrived at the following design principles for FuturICT.[23]

### 6.1 Compliance with Existing EU Data Protection Directives and Privacy Requirements

An obvious prerequisite is that FuturICT has to comply with existing data protection directives and privacy requirements within Europe. For example, all information that refer to individuals will be anonymized to satisfy the extant EU Directive 95/46/EC. All guidelines and ethics codes for social science research of the Academies of Science of the FuturICT member states, of the Academia Europaea, and of the European Science Foundation will be applied. Also the outcomes of large and relevant EU ICT & privacy projects such as *Future of Identity in the Information Society* (FIDIS), *Privacy and Identity Management for Europe* (PRIME), *Privacy Incorporated Software Agent* (PISA), *Ethical Issues of Emerging ICT Applications* (ETICA), *Gentle User Interfaces for Elderly People* (GUIDE), or *Data without Boundaries* (DwB) will be incorporated in the ethical framework. Relevant reports of the Information and Communication Technologies Advisory Group (ISTAG) and the European Group of Ethics (EGE) will also be utilized.

Considerable research efforts will go into the analysis of privacy and data protection in social E-science research. Also the development of new methods to mine data that take privacy issues into account better than is currently done in science and business is an important research topic addressed by FuturICT. New ways of privacy-protecting data storage and mining will be developed and tested. The proposers may thus consider elaborating new policies and procedures to establish future standards and adequate protection of privacy within CRPs involving massive data mining.

---

[23] Some of the principles and methods mentioned are outlined in [22], Section 8.

### 6.2 Enabling Informed Consent

When using the internet, people disclose various kinds of personal information in various ways. It cannot be assumed that disclosure of personal data on the internet is the result of a truly voluntarily and deliberate choice. However, in research projects, voluntarily participation is considered a basic ethical requirement. This basic assumption overlaps considerably with the principle of informed consent, as required by the European Data Protection Directive, which gives individuals an right of control over personal information. There is no unanimous definition for informed consent, but according to Diener and Crandall [12] it is "the procedure in which individuals choose whether to participate in an investigation after being informed of the facts that would be likely to influence their decision". In principle, any decision can be considered as implying informed consent if it has been taken after being provided with the amount of information that a reasonable and prudent person would want to have. In the internet, this condition is seldom the case. In fact, it is both possible and relatively common for individuals to access Web sites without reading the terms and conditions (which usually is a long and abstract text). It is also unlikely that most people would understand the full contract, while they actually have to approve it in order to get the requested service. Rejecting them at the cost of no service does not give users a reasonable choice. Under these circumstances, people may nominally give consent, but without being fully aware of or agreeing with the terms and conditions. Such a situation would not be considered as informed consent.

This contravenes a widely accepted principle in social science ethics that states that "as far as possible, participation in sociological research should be based on the freely given informed consent of those studied"[24]. Moreover, fully informing the respondents it is not yet enough, since researchers should endeavor to make sure that the participants of an experiment have fully understood risks and consequences (this applies in particular for physically or mentally challenged individuals). The Privacy by Design approach aims at producing mechanisms, tools and applications that express respect for autonomy, should enhance informed consent, and would allow the user and citizen to think of the information practices in which she takes part as "hers".

### 6.3 Deliberate Participation

The simplest possibility to do social data mining is to do it with data that individuals share or make publicly available knowingly and deliberately. For example, some Web sites – such as Blippy.com, Skimble.com or Swipely.com – collect everything from consumer data over the last movie you have seen up to how many push-ups you have done in your last training session. Participants of these Web services intentionally make their data available to everybody, and they can be analyzed in any possible way. The only concern from a statistical point of view is that the set of people participating in these Web 2.0 activities is not representative for the whole population, i.e. one would need to make complementary analyzes in order to learn, how it is possible to correct for biases in these data. Typically, participants are younger than average and are not concerned to share their data.

Further data can, in principle, be analyzed by crawling the Web by automated programs (e.g., "spiders"). This data consists of electronic traces of, e.g., shopping activities at eBusiness platforms or social networking activities. They are accessible to everybody in small numbers. It is not clear whether and how much people care about the possibility that a company or a scientist analyzes these data in large amounts, although there are certainly problematic applications of this kind, e.g.. when the resulting data sets are sold to third parties with unknown intentions. In the project *Gaydar* (made public by the *Boston Globe* in September 20, 2009), the MIT demonstrated recently how easy it is to filter out from publicly available data sensitive personal information, which may be misused. This study predicted the sexual orientation of Facebook users by analyzing the publicly accessible pictures of their friends. As the recent discussions about the activities of large data mining companies demonstrate, legal regulations against unauthorized processing of individual data are urgently required. Scientific studies, which lead to discoveries of public interest, may have a better justification, but it must nevertheless be decided in each single case, whether individual rights are violated and what the public benefit is of such studies. Sheer curiosity and the publication of a scientific paper may not be a sufficient justification, and therefore, the consultation of an ethical committee seems appropriate.

---

[24] Paul Spicker, Social Research Update, see: http://sru.soc.surrey.ac.uk/SRU51.pdf

As a consequence, it would be much better to work with data that people provide intentionally for a given purpose. Statistical samples can already be quite useful. Special "on-demand-data-gathering" tools could allow people to easily opt-in and opt-out of data-collection programs in a situation-specific way. For example, while people may usually object to provide their data, it is likely that the participation rate increases in special situations such as crises, where people tend to change their priorities and make a contribution. However, it is fundamental that the gathered data will be used only for the purpose people have explicitly given consent to and that the processing of data should be allowed only for a certain time period. For sensitive data mining activities it would be appropriate to apply the standards followed in clinical studies today. In order to support on-demand participation, particular trust-worthy internet platforms should enable the case-wise sharing of personal data according to the specified purposes.

**6.4 Obfuscation, Anonymization, Surrogacy, and Randomization**

A standard practice to protect privacy, which is also included in the data protection directive 95/46/EC, is to transform the source data into an anonymous or obfuscated version with a quantifiable privacy guarantee (e.g., the probability of re-identification), but that still allows answering the analytical questions correctly when using the transformed data, within a quantifiable approximation that specifies the data utility. This process is known as data anonymization, a.k.a. statistical disclosure control (see [24]). In this way, data containing private information can be managed when performing research projects, in order to prevent any illegal use or dissemination of the information or any abuse by third parties. Scientists need to take appropriate measures to ensure the best practice of managing such kind of information.

The problem here is, that anonymization may not guarantee that the identity of individuals cannot be revealed. Substantial research has been and is currently being performed in the database community on privacy preserving data mining, reflecting the importance of this subject (for a comprehensive state-of-the-art summary see the *Privacy-Preserving Data Publishing Survey*, [17]). Research in privacy-preserving data analyzes has produced methods and tools aimed at publishing data under a privacy-preserving shield. For example, data are made anonymous with respect to a certified trustable anonymity notion, which essentially guarantees that the probability of tracing back any data to the identity of the person to whom the data originally belongs is so low that it can be considered null in practice. Another active research line concerns the privacy issues in case of mobility data such as those produced by location aware devices [1,19].

Nevertheless, there are still a number of open problems, and many approaches are standing next to each other, lacking user-friendliness, integration, and a consequent systemic approach. Problems occur in particular when data sets contain a list of many different features, and some combinations of features are rare. As a consequence, such data must be sufficiently coarse-grained (see Section 6.6) and/or randomized to make sure that combinations of features occur in sufficiently large numbers and cannot be individually resolved. Furthermore, it must be avoided to save lists with many features in one single data set. It is safer to store them separately on different computers and to access the separate data sets only with programs, which are guaranteed to determine coarse-grained properties only such as (sufficiently rough) statistical distributions. The resulting derivative data sets should be comparatively small and unspecific.

Data with sensitive information allowing e.g. to identify infrastructure targets vulnerable for terror attacks should not be made publicly available for security reasons. However, it may still be in the public interest to provide surrogate versions of the data sets to researchers, which would have to be prepared by experienced and authorized research groups for public research. In surrogate data sets, the relevant statistical properties are the same as in the original set, but the underlying individuals (persons, companies, etc.) are randomly reshuffled and not identifiable anymore.

The generation of anonymized and surrogate data sets should be done by particularly qualified and trustable institutions, while a larger number of people can work with the resulting, less critical data sets. To protect the original data sets from theft and unauthorized access, the specially secured and authorized data centers should store them in an encrypted way, and decryption should be done only piecewise and for the milliseconds, when the derivative data are generated. All commands and source codes of computer programs involved in sensitive operations should be automatically protocoled on a separate server, which is inaccessible to persons who are not authorized to deal with original data sets. Such an institution may also be authorized to remotely deactivate software involved in processing information that could be attributed to persons. An ethical committee should accompany and supervise these terminating activities.

### 6.5 Coarse-Graining, Hierarchical Sampling, and Recommender Systems

As indicated above, in case of sensitive personal data (such as religious affiliation, diseases, or sexual orientation), it must be ensured that individuals and group memberships cannot be identified from socio-economic data sets. For this reason, data sets for statistical analyzes must be coarse-grained in a suitable way [24]. This may also be done in the course of real-time data mining approaches, if they are suitably designed. For example, to determine congestion on a freeway, it is possible to analyze mobile phone usage data, but it is not necessary to know who is calling whom and what the content of the call is. The same applies to GPS localization information of mobile phones, if, for example, the distribution of people is determined for the sake of an efficient evacuation. I these cases it is necessary to ensure that any potentially sensitive data (such as the underlying phone number) is deleted before the statistical evaluation is performed. However, as the case (made public in 2010) of Wi-Fi recordings by Google Street View cars has shown, transparency is needed for such applications, as one needs to make sure that really no sensitive data are stored. In principle, it could be legally required that the underlying algorithms are published, and no algorithms may be used which are not open source.

One particular approach in reality-mining could be a hierarchical sampling via ad-hoc networks of, for example, sensors or mobile phones, where detailed information is only processed locally, and any transmitted information undergoes a certain level of aggregation. That is, as data are distributed over larger distances, they undergo several aggregation steps, which may be imagined like a hierarchical sampling method. Whoever wants to analyze a large data set, would only get a coarse-grained view of the data that is appropriate for the purpose of the analysis. Whoever managed to see data on a lower and, therefore, detailed level, would only have a very short-sighted and limited view, i.e. see very little. It appears, however, that the technical details of such systems matter in order to be sufficiently privacy-protecting and acceptable to users of the resulting services (e.g. location-based ones). A transparency of the data processing algorithms and related legal regulations appears to be needed. It should be explicitly forbidden and prevented to collect and store low-aggregation-level data. It must be ensured that they are deleted directly after they have been processed and before they are transmitted. To be widely acceptable, the processing should take place in the technical devices used by the individuals, and not on company-owned infrastructures (as it is common today).

A possibility to make low-level data robust to interception could be conceived of along the following lines: given that the data of interest can be represented as points in a (quasi-)continuous space, one could add random numbers according to a certain statistical distribution. Rather than transmitting the correct value (such as the exact location of the individual), a random number ("noise") would be added, before the value is transmitted to the ad-hoc network performing the reality mining. Such random falsification would make low-level-aggregated data useless and create a "foggy" situation that protects the individual from being revealed. However, the aggregation of the individual data could still lead to reasonably accurate results, as errors average out in a statistical sense.

Services of recommender systems, of course, need to target an individual specifically, which seems incompatible with overlaying noise. However, recommender systems could still be realized by applying a two-component strategy: The first component would be a rough search, which does not consider individual information or preferences. Among the search results, the personal computer or smart-phone of the user would then select the individually fitting search hits, products, or advertisements, based on personal information and preferences that are exclusively stored on the individual computer. Thus, recommender systems should not follow a push approach, where individually customized recommendations are transmitted to the user, but a pull-approach, where the user selects in confidence one option out of a larger spectrum of recommendations in a way that does not reveal his or her preferences. Individuals who are even concerned about storing personal information and preference data on their own computational device should have the possibility to turn off the second component, which would then result in untargeted research results and in recommendations, which would not be individually customized. The same approach can be used in connection with location-based services, the great comfort of which many people do not want to miss anymore.

### 6.6 Pseudonyms and Virtual Identities

Another instrument to protect privacy refers to pseudonyms and virtual identities.[25] This may be useful when, e.g. studying social interactions using multi-player on-line games such as Second Life; a research technique which is used as a complement to lab and Web experiments. The advantage of these games with respect to privacy is that players can participate under pseudonyms, without revealing their real identity.

---

[25] Interesting work in that respect has been done by IBM's Zuerich based privacy lab. See for example the work of Jan Camenisch: `https://researcher.ibm.com/064446t56x15777w`

From an experimental point of view, this has some side effects, as people may behave differently under anonymous conditions as compared to conditions with face-to-face interactions; although there are a number of behaviors, which come out quite realistically also under these conditions. Artifacts when studying multi-player on-line games result from the following facts (we assume that the system does not allow the registration of several identical pseudonyms):

1. People may change identities, i.e. register as a new user if their previous behavior is sanctioned by other players or by the system ("whitewashing").
2. People may use multiple identities, potentially also in parallel.
3. People may buy an identity (pseudonym) with high reputation or scores from somebody else

   To overcome these problems, the following measures can be taken:

- Everybody could get a unique virtual identity, which would be needed to create unique pseudonyms.
- Registering a new identity could be made very time-consuming or costly.
- People may be allowed to join a multi-player on-line community by invitation only (and there would be separate lists of members and pseudonyms, which would be secret and encrypted).
- One can perform behavioral consistency checks to reveal the use of the same identity by different people, and the matching of pseudonyms with the unique virtual identity could be sporadically checked (by requiring to enter it).

Pseudonyms and virtual identities are, however, not only used in games, but also in real-world applications, raising similar issues. In particular, the uniqueness of such an identity has to be guaranteed. A unique virtual identity can be generated by a trustable public institution such as the registration office. This is practically an electronic signature that can be used to submit documents such as tax declarations or payments. Note that there are already private companies offering trusted virtual identities/electronic signatures, among which Verisign, GeoTrust and Thawte. The unique virtual identity would have a finite validity (i.e. it would have to be regularly renewed), and plausibility checks for identity thefts would have to be made, to invalidate stolen identities (such as for credit or debit cards). To reveal the real identity behind a virtual identity in case of a severe crime, this should require the simultaneous agreement of several independent authorities.

### 6.7 Anonymous Lab Experiments

Social behavior can also be studied in lab experiments. In these experiments, it is important to be able to ensure anonymity of the participants, as they may otherwise not reveal their true opinions or their normal behaviors. This can be made by ensuring that the experimental subject does not meet the experimenter, and maybe not even other experimental subjects. There are different ways of implementing such a design (for more details, see [22]). Particulary suitable in that respect are Web experiments, for which experimental subjects could be recruited in different ways (posters in public areas, calling for participation at a specified time via a certain Web page, advertisement on a heavily frequented Web portal, etc.). Here, a large number of people would be informed that the experiment takes place at a certain time and they could log on with pseudonyms. The computer would then randomly match individuals to form experimental groups. At the end of the experiment, each individual would get a voucher with a unique code, which can be exchanged for the compensation for participating in the experiment (e.g. from an independent cashier). In doing such experiments, one should also implement methods to avoid that subjects participate multiple times in the same experiment.

### 6.8 Two Sided Transparency

The current generation of global online information sources, embodied by large web search corporations such as Google and Yahoo!, are characterized by two-sided opacity between the information retrieval system and its users. On one side, the methods used to answer the users' search requests are not publicly known, and it is not possible to assess the technical, cultural, or commercial biases that the information retrieval system may introduce. On the other side, the specification of users' identity and of the purpose of their queries is not required – albeit that web search corporations extensively use profiling techniques to understand motivations and interests of their users. Two-sided opacity implies two-sided (mutual) irresponsibility: either party is not accountable for the quality and purpose of their interactions and exchange. In a trusted ICT ecosystem with mutual responsibility and mutual exposure, we may create more robust ways to link our real identity to our virtual identity in the system or network environment. This would mean a shift from 'two sided opacity' to 'two sided transparency'.

### 6.9 Incentivizing Mutual Data Protection: Coprivacy By Design

Another promising approach that will be explored and developed in FuturICT is "coprivacy by design". This approach revolves around finding incentives for privacy respecting behavior. This *quid pro quo* approach, which relies on a selfish interest in mutual collaboration in data protection, has been formalized by Domingo-Ferrer [13] in the notion of coprivacy. A protocol or an interaction is coprivate if the best option for a player to preserve her privacy is to help another player in preserving his privacy. Coprivacy makes an individual's privacy preservation a goal that rationally interests other individuals: it is a matter of helping oneself by helping someone else. Coprivacy can be described in terms of Nash equilibria and the following generalizations and extensions of it can be developed: i) general coprivacy, where a helping player's utility (i.e. interest) may include earning functionality and security in addition to privacy; ii) mixed coprivacy, where mixed strategies and mixed Nash equilibria are allowed with some restrictions; iii) correlated coprivacy, in which Nash equilibria are replaced by correlated equilibria. Coprivacy can be applied to any peer-to-peer (P2P) protocol. In [13] coprivacy-by-design is illustrated in several application areas: P2P anonymous keyword search, content privacy preservation in social networks, vehicular network communications, and controlled content distribution and digital oblivion enforcement.

### 6.10 Compliance with a Code of Conduct

In order to increase the awareness of the various ethical issues that may arise in the progress of FuturICT, creating a "code of conduct" could be a suitable instrument. A code of conduct relies on a variety of ethical principles that seem acceptable to all scientists working in this field, such as:

– promote human well-being,
– increase the self-awareness of society,
– reduce vulnerability and risk in society,

- increase resilience (the ability to absorb societal, economic, or environmental shocks),
- avoid loss of control (sudden, large and unexpected systemic shifts),
- develop contingency plans,
- explore options for future challenges and opportunities,
- increase sustainability,
- facilitate flexible adaptation,
- promote fairness,
- increase social capital and the happiness of people,
- support social, economic and political inclusion and participation,
- balance between central and peripheral (global and local) control,
- protect privacy and other human rights, pluralism and socio-bio-diversity,
- support collaborative forms of competition and vice versa ("coopetition").

This unstructured list certainly is not yet the envisaged code of conduct; rather, it provides a starting point. Although creating such a code is not sufficient to ensure ethical research within FuturICT, we consider its establishment as an important aspect in increasing the awareness for ethical issues.

A major objective of such a code (and the other design principles mentioned in Section 6) is the prevention of unethical use of powerful ICT tools. We should require a minimum level of qualification and training, including ethical issues, from people working with them. The activities of these ICT systems may need to be recorded and monitored in a way that is sufficiently transparent; security measures will depend on the size of the simulations and application areas. Finally, the public should be informed about relevant activities in academia, government and private companies in that respect.

# 7 Other Ethical Concerns

Privacy is a core ethical issue within FuturICT, but not the only ethical question connected with large-scale data mining activities and their application to the social world. Many of those problems are ambiguous and prone to inevitable disagreement, i.e. the correct answer cannot be deduced from general rules, principles or norms. Thus, we prefer an ethical approach outlined in the EU ETICA project[26], namely a pluralistic, pragmatic and applied ethics of ICT that is not bound to a particular ethical tradition such as Utilitarianism, Kantianism, Virtue Ethics and their modern versions. As the modern *conditio humana* is characterized by value plurality, the moral analyzes in the context of FuturICT should accommodate a methodological and theoretical pluralism [45].

In the following, we raise a number of open ethical questions, to which, of course, we cannot provide definite answers here. In Section 8, we outline the institutional design of FuturICT that should allow us to address these questions.

### 7.1 Social Goods, Equality and Fairness

Many private companies have access to a wealth of data that may help to shed light upon public policy problems and their solutions. Wall Mart, Google and Facebook, are able to outperform in some predictive and explanatory tasks governmental institutions, e.g. with respect to Public Health. This leads to the paradox that societies and their individual members are generating data about themselves that society itself often cannot use to improve itself or the quality of life of its citizens. The more powerful large-scale data mining activities are in prediction, the more pressing will become questions like: Who will use this knowledge (governments, companies, the public)? For which purposes? Would there be a moral imperative to make this knowledge available to stakeholders who are not able to generate it by themselves (e.g., developing countries, NGOs)?

---

[26] See deliverable 2.2. of the *Ethical Issues of Emerging ICT Applications* project, accessible through: www.etica-project.eu

FuturICT seeks to address this situation – and an important step to that end is to construe social knowledge as a public good, a social good or a merit good [7]. As starting point, we rely on the notion of 'social primary goods' according to Rawls [36]. Those can be characterized as goods that anyone would want regardless of whatever else they wanted. They are means (or resources, broadly conceived) that have a use independent of someone's plan of life. The examples that Rawls give include basic liberties (freedom of thought, liberty of conscience, etc.) as well as rights required for the pursuit of final ends (freedom of movement, free choice of occupation, etc.), powers and prerogatives of offices of responsibility, income and wealth (broadly understood), and the social basis of self-respect.

We propose that information pertaining, e.g., to health, politics, employment, education, food quality, transport and mobility, safety and security, housing, economic prospects, or weather conditions, should be understood as primary social good, too. In doing so, we refer to information types and not to specified information content, whose importance may vary (e.g., if someone has settled, housing information may be less relevant for this individual). But every reasonable and rational person would like to have access to such information types, irrespective of his or her individual life-plan and aims in life. This suggests that access to the information that FuturICT intends to collect qualifies as a social primary good. It also implies that the principles of social and distributive justice suggested by Rawls apply to them and would be reformulated as follows:

1. Everyone must have an equal right to the most extensive set of equal basic (information) liberties consistent with those of all others.
2. Inequalities (in access) – if they cannot be avoided – must be arranged so that
   a) they must be of the greatest benefit to the least-(informationally) advantaged
   b) offices and positions (that bring information or capabilities to use information to one's advantage) must be open to everyone under conditions of free and equal opportunity

To what extent this proposal to understand information as primary social good holds, certainly needs further investigations along at least three lines. First, one has to look to the critics of Rawls's suggestion, most prominently Amartya Sen and Martha Nussbaum, who favor a different perspective on primary goods called the capability approach. Sen has suggested that some individuals may need more than an equal share of the pie to achieve fairness and equality. For example, giving everyone the same amount of rice can be unfair if some are recovering from an illness, are pregnant, or have a higher metabolic rate. Thus, we need to take into account the individual capabilities to convert shares of primary goods into what really matters in human lives. Transferring this into the realm of information: The capability to convert data into something worthwhile (information and practically relevant knowledge) for themselves is crucial for the idea of equality in the information realm. So this suggests that FuturICT will have to accommodate a broad range of users with different technical and cognitive skills in order to realize effective equal access.

Second, based on the objections by Michael Walzer and others (see Section 4.3), one can argue that the "information principles" of social and distributive justice will need to take on different forms depending on the sector or sphere of society to which they apply. Criminal justice data, commercial data, health care data, will each have rules and principles of access that are appropriate and specific to them. So, although Rawls conception of a fair society seems to provide a sound moral default position, it need to be elaborated further along the lines of Sen's and Walzer's critique.

Third, one has to take into account that information is positional, i.e. the value of a certain information type is a function of its ranking in desirability by others. Rawls' principles have difficulty dealing with positional goods. This applies especially to the difference principle (2a). This principle would imply, in the context of FuturICT, that certain differences or inequalities in access to data would be justified only if they would indirectly advantage those who are worst off in terms of access and utilization of data. But there is a problem with this application of the principle to informational goods. Consider the following distributions A, B, and C of shares of goods (represented by numbers) over three positions in a group of three individuals (p,q,r). A: (15,10,5); B: (6,5,4); C: (3,3,3). The difference principle would favor the very unequal distribution A on the basis that the worst off – those who get 5 – are better off under A than under distributions B and C, where they would only get resp. 4 or 3. This applies to non-positional goods such as apples, but not to positional goods such as data or weapons. If we all have guns and the distributions A, B, C refer to the number of bullets each person gets, it is clear that the most egalitarian distribution (C) is the preferred option. If the numbers represent, e.g., access to financial databases, distribution A would leave the worst-off in a comparatively poor information position. Although the worst-off person under A do better than the worst-off under B and C, the person who does best under A has three times as much information at his or her disposal. The difference principle allows for differences that really make a difference.

Furthermore, vertical positionality, as described above, needs to be distinguished from horizontal positionality. A good is horizontally positional if its value depends on whether other goods are available (to oneself or others). The first part of the code of a safe is of value only if I or some else has the second part. The first part has a specific value to me only on condition that the other part is in relevant ways available to me or to others. Such relations between information types also have to be taken into account.

The upshot of this discussion is that there is a *prima facie* obligation regarding FuturICT to reduce vertical positionality of information (the extent to which some people can exploit a head start in access at the expense of others) and that the architecture (data and models) ought to be structured in such a way as to perspicuously represent horizontal positionality and present relevant meta-data that indicate important value-adding relationships.

### 7.2 Ownership

Who will own the algorithms and the outcomes of the data mining activities achieved in the FuturICT context? Intellectual property is often discussed in terms of ownership of data used for input, but an equally important question would seem to be: Who owns the predictions and the fruits of the analyzes? As they could potentially be subject to patent protection for computer programs and business methods, a rigorous analysis of the implications of intellectual property protection for data mining activities is needed.

Given the account of information as a primary social good it seems that FuturICT should not be construed as a proprietary environment, but would have more similarities with open source, open knowledge and creative commons environments and commons based peer production. To what extent this is actually compatible with the various traditions and practices with respect to intellectual properties at the member countries and host institutions of FuturICT will be an important field of study.

### 7.3 Reliability of Information and Quality of Data

Another problem for FuturICT is how we should deal with competing claims of systems, models and simulations? If an early warning system recommends certain activities, how should one respond to such recommendations? For example, how to handle situations, in which a scarcity of resources occurs? If policy is based on predictions, how open is the system to critical review? Who will know and understand the algorithms? How can mistakes in algorithms be identified and rectified? It seems very obvious to construe data quality policies and monitoring analogous to best practices in fields where data quality is all important (health, tax, customs, safety).

This point closely relates to another issue of ethical relevance, namely how to ensure that the results provided by FuturICT are actually used by decision makers and players. As mentioned in the introduction, already now relevant knowledge for, e.g., preventing crowd disasters, is available – but it is not applied in critical situations. Within FuturICT, one therefore has to take into account that "delivering" the results will not be sufficient, one has also to solve a 'transfer task' such that the knowledge gained also becomes of practical use.

### 7.4 Liability for Use and Misuse of Results

The rationale of FuturICT is that it will be of help in designing interventions in the context of dealing with social problems. This raises the question of who will be liable for damages if the information that is acted upon proves false or inaccurate. It has to be noted that the users of FuturICT are epistemically dependent on it. The information residing in FuturICT is supposed to be highly relevant to thinking about urgent and important social questions. It is most likely the most authoritative repository of data and models relevant to answering these social questions. This makes the designer and managers responsible in a special way.

This responsibility also extends to another issue. As large-scale data mining activities are increasingly successful in predicting, they will constitute an extremely powerful tool – and thus may be misused. For example, if a 'crisis observatory' detects increasing risks for a stock market crash, making this knowledge available to selected marked players would certainly qualify as a misuse. One therefore has to be alert for such possibilities within the organizational design of FuturICT.

## 8 Governance and Institutional Design

Any research project implemented within the FuturICT program that addresses topics such as reality mining can be potentially harmful for individual privacy or be instrumental in wronging individuals or groups in a range of ways. Furthermore, various other ethical issues may arise, as described in the previous Section. The management and Scientific Committee [41] should therefore pay all due attention to potential problems related to data mining, privacy, and proprietary issues.

The proposals for LEP/Explanatory/CALLS therefore include concrete prevention mechanisms and Privacy by Design solutions. They explain in detail how the consortium will address ethical issues in each particular case. Privacy Impact Analysis (PIA) and a Responsible Research and Innovation Impact Assessment along the lines developed by the European Commission[27] need to be conducted well before the launch of the project, and need to be updated on a regular basis. The involved Explanatories, institutions and projects should rely on an Ethical Committee, which will be actively involved when and where appropriate. This Committee gives direction and supervises the activities. Collaboration in that respect will be sought with the European Data Commissioner and the Chairman of the Article 29 Working Group for consultation and representation in the Ethical Board of FuturICT. The community of FuturICT researchers should all be known and identified, log files of interactions and activity in the system will be kept and be open to periodic review. There will be a societal panel where complaints can be filed.

---

[27] See: http://ec.europa.eu/research/science-society/document_library/pdf_06/rri-report-hilary-sutcliffe_en.pdf

In particular, the following instruments will be included into the institutional design of FuturICT:

- *Commons Based Peer Production and Social Computing:* The sociologist Robert Merton has suggested the following basic norms for scientific activity: Communalism, universalism, disinterestedness, organized skepticism (CUDOS) [27]. The Academies of Social Sciences and the Academia Europaea have over the last years built on these and other broadly shared values for social scientists. FuturICT will take the fruits of these discussions as a point of departure for the norms of the scientific community within FuturICT. In addition the ethos of commons based peer production [3] is seen as a source of inspiration.
- *FuturICT License Agreement:* The community of users will form a community on the basis of an agreement (the relations with the Common License Model will be explored). This agreement stipulates that results will be given back to FuturICT, that they cannot be commercially exploited, that they can be shared with others on the same conditions under which data and models where acquired and utilized, and that their usage needs to be in compliance with all relevant principles and rules that have been established by the FuturICT consortium in close collaboration with external stakeholders and relevant legal bodies.
- *User Panel:* Users of FuturICT will be represented in a user panel that can deal with issues, complaints and questions of users, stakeholders, citizens, organizations. Full use will be made of the extensive experience with user involvement, participatory design, usability studies, in order to make user experience, interests and feedback effective.
- *Upstream Engagement:* Already in the early stages of FuturICT the media were very interested in this project. Its outreach, public engagement and popularization of the aims, as well as scope, possible results and problems drew public attention. FuturICT will reach out and disseminate results and news to a large audience.
- *Ethical Board:* Since it is impossible to analyze all ethical, social and legal aspects, issues and problems concerning FuturICT in advance, an ethical board will be established and a substantial ethics program will be implemented. Ethics research will be a substantial part of the FuturICT FET Flagship and will explore new issues as they arise. This line of research and the board will benefit from recent EU projects in ethics of emerging ICT (see Section 6.1) which have been built up over the last decades.

## 9 Conclusions

1. FuturICT is a community of scientists and users and citizens that is ready to jointly take the responsibility to shape a global knowledge resource that will provide on-line functionality and social networking platforms, that will both lead to deeper individual and collective understanding of modern society and its problems. It will do so in an open and accessible way and construe the knowledge it opens up as a public good. This goal is a moral imperative, because the available big data concerning human activities give us a social microscope that allows us to see aspects of our collective behavior that were not visible before – a promise of new knowledge that we cannot afford to ignore. At the same time, this goal has formidable moral repercussions, because the findings envisaged here have the potential of uncovering some of the mechanisms of our society that were hidden thus far. This knowledge may ultimately change our societies profoundly – for the good or the evil. This is why we ought to pursue FuturICT, and insist that our shared values need to inform its foundations.

FuturICT will develop a Global Participatory Platform, which will allow people to play a crucial role in the articulation and identification of our shared values and take part in producing and consuming of collective knowledge. In the participatory platform, high levels of transparency, protection and awareness will be achieved:

i) Data liberation: new means for supporting each participant's full control of own personal data / knowledge, moving from 'informed consent' to 'awareness'. Data liberation means also the right (and tools) to withdraw personal data at any moment in full; the right/tools to oblivion (having personal data forgotten after a specified amount of time); the right/tools to have full access to the collective knowledge as a public good; the right/tools to know when and what personal data are used in modeling, profiling or decision making.

ii) Authenticated anonymity as a tool for trusted participation through a form of pseudonymity which combines the merits of authentication (trust, mutual responsibility, awareness, resilience to hacking) and anonymity (protection of private sphere, liberty, transparency).

2. In the second, mature, phase of FuturICT, its trusted participation platform should provide valid incentives to opt-in for many: the greater protection/trust level and the possibility to create knowledge services that are not feasible without collective participation. Two metaphors to sustain why FuturICT might be successful, today, despite its very ambitious goals:

i) The "city" metaphor: the reason to abandon tribal or nomadic life and form villages and cities was probably based on the trade-off between an acceptable reduction of liberty and privacy and a more prosperous, trustable and safe place to live. Many people today seem to accept the transition from a wild ICT ecosystem towards a trusted participatory ICT ecosystem that amplifies everybody's knowledge while respecting shared values.

ii) The "election" metaphor: participating in FuturICT has the chance of becoming somewhat similar to casting your vote in a mature democracy. Casting your vote is like sharing your data: in itself, your individual vote is marginal: it is very unlikely that it will decide the elections, as well as your personal data will unlikely affect a global model or simulation. But, at the same time, your vote and your piece of data – your participation – is essential to shape society, and to shape everybody's knowledge about society.

3. Other programs or global corporations may share, to some extent, some of the goals of FuturICT. Some organizations aim to develop global sensor networks to monitor environmental phenomena, and the major web corporation declares its mission as "organizing the world's information". But only FuturICT is set to create a global knowledge resource as a public good for everyone, based on the shared values of liberty, justice and moral responsibility.

*The publication of this work was partially supported by the European Union's Seventh Framework Programme (FP7/2007-2013) under grant agreement no.284709, a Coordination and Support Action in the Information and Communication Technologies activity area ('FuturICT' FET Flagship Pilot Project).*